\documentclass[twocolumn,english,aps,prb]{revtex4}
\usepackage[T1]{fontenc}
\usepackage[latin9]{inputenc}
\usepackage{float}
\usepackage{amsmath}
\usepackage{graphicx}
\usepackage{amssymb}

\makeatletter
\@ifundefined{textcolor}{}
{%
 \definecolor{BLACK}{gray}{0}
 \definecolor{WHITE}{gray}{1}
 \definecolor{RED}{rgb}{1,0,0}
 \definecolor{GREEN}{rgb}{0,1,0}
 \definecolor{BLUE}{rgb}{0,0,1}
 \definecolor{CYAN}{cmyk}{1,0,0,0}
 \definecolor{MAGENTA}{cmyk}{0,1,0,0}
 \definecolor{YELLOW}{cmyk}{0,0,1,0}
 }

\makeatother

\makeatother

\usepackage{babel}

\begin{document}

\title{Entanglement in the symmetric sector of $n$ qubits}

\author{P. Ribeiro}

\email{ribeiro@cfif.ist.utl.pt}

\affiliation{CFIF, Instituto Superior Técnico, Universidade Técnica de Lisboa,
Av. Rovisco Pais, 1049-001 Lisboa, Portugal}

\author{R. Mosseri}

\email{remy.mosseri@upmc.fr}

\affiliation{Laboratoire de Physique Théorique de la Matière Condensée, CNRS UMR
7600, Université Pierre et Marie Curie, 4 Place Jussieu, 75252 Paris
Cedex 05, France}


\begin{abstract}
We discuss the entanglement properties of symmetric states of $n$
qubits. The Majorana representation maps a generic such state into
a system of $n$ points on a sphere. Entanglement invariants, either
under local unitaries (LU) or stochastic local operations and classical
communication (SLOCC), can then be addressed in terms of the relative
positions of the Majorana points. In the LU case, an over complete
set of invariants can be built from the inner product of the radial
vectors pointing to these points; this is detailed for the well documented
three-qubits case. In the SLOCC case, cross ratio of related Möbius
transformations are shown to play a central role, examplified here
for four qubits. Finally, as a side result, we also analyze the manifold
of maximally entangled 3 qubit state, both in the symmetric and generic
case. \\

\end{abstract}
\maketitle
\global\long\def\ket#1{\left| #1\right\rangle }

\global\long\def\bra#1{\left\langle #1 \right|}

\global\long\def\kket#1{\left\Vert #1\right\rangle }

\global\long\def\bbra#1{\left\langle #1\right\Vert }

\global\long\def\braket#1#2{\left\langle #1\right. \left| #2 \right\rangle }

\global\long\def\bbrakket#1#2{\left\langle #1\right. \left\Vert #2\right\rangle }

\global\long\def\av#1{\left\langle #1 \right\rangle }

\global\long\def\tr{\text{Tr}}

\global\long\def\im{\text{Im}}

\global\long\def\re{\text{Re}}

\global\long\def\sign{\text{sgn}}

\global\long\def\abs#1{\left|#1\right|}

The potential power of quantum information, either for cryptography
and computation purpose, is largely based on the subtle concept of
quantum entanglement \cite{Chuang_2000}. In a system composed of
$n$ two-level entities (qubits), a generic state is entangled, \textit{e-g}
it cannot be written as a separable product of states belonging to
each constitutive part. While it is rather easy to characterize entanglement
for a 2-qubits system, the task of quantifying the amount of entanglement
carried by the total system is very difficult, for increasing $n$.

Several entanglement measures have nevertheless been proposed (see
\cite{Guhne_2009,Plenio_2007} for a comprehensive reviews), and their
behavior under state transformation studied. Important cases are given
by those quantities which remain invariant under (stochastic) local
operations and classical communication, noted (S)LOCC \cite{Bennett_2000,Dur_2000}.
Stated as operations performed in the multiquibit Hilbert space $\mathcal{H}$,
the latter read $\otimes_{i}M_{i}$, called local unitaries (LU) for
LOCC (with $M_{i}$ a unitary matrix), and invertible local operations
(ILO) for SLOCC ($M_{i}$ a matrix with non vanishing determinant). 

One aims to find a complete set of such invariants that parameterizes
the orbit space $\mathcal{H}/\otimes_{i}M_{i}$. Physically this means
that states can only be obtained from each other with a local transformation
(LU or ILO) if they share the same set of invariants. In the LOCC
case, LU invariants can in principal be written as polynomial functions
of the state components \cite{Grassl_1998,Luque_2003,Levay_2005}.
However their number proliferates with $n,$ and finding explicit
expressions becomes challenging; moreover their physical relevance
is not necessarily obvious. Upon enlarging the set of operations that
can be performed locally, like passing from LU to ILO, the number
of entanglement classes can generally be reduced. 

In this paper we consider symmetric $n$-qubit states, and analyze
their entanglement properties under LOCC and SLOCC. Such states have
been the subject of several recent studies \cite{Hubener_2009}\cite{Li_2009}\cite{Chen_2010}\cite{Markham_2010}\cite{Martin_2010},
with even some experimental \cite{Wieczorek_2008} realizations or
proposals \cite{Bastin_2009_b}. In that case most of the relevant
bipartite entanglement criteria was shown to coincide \cite{Toth_2009_b}
and generic entanglement measures usually simplify.

We use the Majorana representation \cite{Majorana_1932}, which characterizes
such a state as a collection of $n$ points on a sphere, and derive
the entanglement invariants in terms of the points arrangement. In
the LOCC case, invariants can indeed be built from the inner product
of the radial vectors pointing to these points; we explicitly derive
the well known 6 LU invariants for three qubits. In the SLOCC case,
we show how sets of cross-ratio, invariants under ILO related Möbius
transformations, play a central and clarifying role, and relates to
a recently proposed classification of entanglement classes \cite{Bastin_2009}.
For four qubits, the most generic SLOCC invariant is simply related
to the Klein modular invariant \cite{Klein_1921}. Finally, and as
a side result, we also precise the manifold of maximally entangled
3 qubits GHZ-like states, both in the symmetric and the generic cases.

\begin{center}
\textit{Majorana Representation in the symmetric sector }
\par\end{center}

The $n$-qubits Hilbert space decomposes into subspaces of constant
total spin $S^{2}=\mathbf{S}.\mathbf{S}$ (where $S=\frac{1}{2}\sum_{i=1}^{n}\boldsymbol{\sigma}_{i}$).
The subspace of maximal spin, $S^{2}=s(s+1)$ with $n=2s$, which
appears once in this decomposition, corresponds to the fully symmetric
sector, spanned by the Dicke basis ($S_{z}\ket{s,m}=m\ket{s,m}$).
Using spin coherent states $\ket{\alpha}=e^{\alpha S_{+}}\ket{s,m=-s}$,
where $S_{\pm}=S_{x}\pm iS_{y}$, any symmetric state $\ket{\Psi}$
can be represented by its Majorana polynomial\begin{eqnarray}
\Psi(\alpha) & = & \braket{\alpha}{\Psi}\\
 & = & \sum_{m=-s}^{s}\sqrt{\frac{(2s)!}{(s-m)!(s+m)!}}\braket{s,m}{\Psi}\alpha^{m+s}.\nonumber \end{eqnarray}
Up to a global unphysical factor, this state is therefore fully characterized
by the set $\{\alpha_{i}\},$ made of the $n$ complex zeroes of $\Psi(\alpha)$,
suitably completed by points at infinity whenever $\left\langle s,s|\Psi\right\rangle $
vanishes: $\Psi(\alpha)\propto\prod_{i=1}^{2s}\left(\alpha-\alpha_{i}\right).$
A nice geometrical representation of $\ket{\Psi}$, by $n$ points
on the unit sphere, is obtained by an inverse stereographic map of
$\{\alpha_{i}\}\to\left\{ \mathbf{v}_{i}\right\} $. The Majorana
high spin spherical representation generalizes (although published
earlier) the spin $1/2$ Bloch sphere; it recently proved quite useful
in the context of collective spin models \cite{Ribeiro_2007}. 

\begin{center}
\textit{Symmetric LU Invariants (SLUI)}
\par\end{center}

A generic local (separable) unitary transformation acting on a set
of $n$ qubits can be written, up to a un-physical phase, in the form
$U=\otimes_{i}e^{\frac{i}{2}\mathbf{h}_{i}.\boldsymbol{\sigma}_{i}}$,
with a collection of vectors $\mathbf{h}_{i=1,..,n}\in\mathbb{R}^{3}$.
In the symmetric sector, we restrict to identical $\mathbf{h}_{i}$,
leading to the simpler form\begin{equation}
U_{s}=e^{i\mathbf{h}.\boldsymbol{S}}.\label{eq:sym_trans}\end{equation}
This implies that, in the symmetric sector, the only states that are
LU equivalent correspond to sets of (unordered) Majorana zeroes which
can be transformed into each other by a global rotation of their representative
vectors $\mathbf{v}_{i}\to\tilde{\mathbf{v}}_{i}=R.\mathbf{v}_{i}$,
with $R$ in $SO(3)$. Moreover, one also expect equivalent entanglement
measures for states that are related by an (anti-unitary) time reversal
operation $T=\otimes_{j=1}^{n}\left(i\sigma_{y}\right)\mathcal{K}$,
where $\mathcal{K}$ is the complex conjugate operator in the computational
basis $\mathcal{K}\left(\sum_{ijk=0,1}t_{i,j,k}\ket{i,j,k}\right)=\left(\sum_{ijk=0,1}\bar{t}_{i,j,k}\ket{i,j,k}\right)$
and $T^{2}=(-1)^{n}$. Geometrically, this corresponds to an inversion
$\mathbf{v}_{i}\to\tilde{\mathbf{v}}_{i}=-\mathbf{v}_{i}$.

An over-complete set of SLUI is obtained from symmetrized products
of the inner-products $v_{ij}=\mathbf{v}_{i}.\mathbf{v}_{j}$, like
for instance with the $c_{k}$ coefficients of $x^{k}$ in the symmetrized
product $\prod_{ij}(x-v_{ij})=\sum c_{k}x^{k}$. It is instructive
to relate them to the standard invariants for two and three qubits.
We make use of density matrices $\rho=\ket{\Psi}\bra{\Psi}$ and eventually
uses their partial trace, with indices in $\rho$ indicating those
parts which have not been traced out. 

\begin{center}
\textit{The two-qubits case}
\par\end{center}

For $2$ qubits, there is one entanglement invariant (if we disregard
the trivial invariant $\tr[\rho]=1$ for a normed state), which we
express here with the single inner product $v_{12}$. It can be given
as the (equal) radius $r_{i}$ of the partial Bloch sphere when tracing
out one of the 2 sub-system. From $r_{i}^{2}=2\tr[\rho_{i}^{2}]-1$,
one gets $r_{i}=\frac{8(v_{12}+1)}{(v_{12}+3)^{2}}.$ Another most
used form is the concurrence \cite{Wootters_1998} running from zero
for a separable state to unity form maximally entangled EPR state.
In the symmetric sector, it reads $C=\frac{4}{v_{12}+3}-1$. Separable
symmetric states corresponds to the case with the two identical Majorana
points , while symmetric EPR corresponds to pairs of antipodal points
($v_{12}=-1$). The latter set is then given by the sphere $S^{2}$
with opposite points identified, the projective plane $RP^{2}$. Note
that a simple but careful analysis, not reproduced here, allows to
extend the EPR case to the full Hilbert space (not only to the symmetric
sector), and recover the well known $RP^{3}\left(\equiv SO(3)\right)$
EPR manifold \cite{Bengtsson_2006}.

\begin{center}
\textit{The three-qubits case}
\par\end{center}

A complete set of six independent LU invariant polynomials is known
\cite{Kempe_1999,Coffman_2000}. For a generic 3 qubit state, \begin{eqnarray*}
I_{1} & = & \tr\left[\rho\right],\\
I_{i=2,3,4} & = & 2\tr\left[\rho_{i-1}^{2}\right]-1,\\
I_{5} & = & \tr\left[3\left(\rho_{1}\otimes\rho_{2}\right).\rho_{12}\right]-\tr\left[\rho_{1}^{3}\right]-\tr\left[\rho_{2}^{3}\right],\\
I_{6} & = & \tau_{3}.\end{eqnarray*}
Again, $I_{1}=1$ for a normed state. $I_{2,3,4}$ are related to
the radius of the (partial) Bloch balls of qubits $\left(1,2,3\right)$
respectively, once the other two are traced out. $I_{5}$ is the Kempe
invariant \cite{Kempe_1999} and $I_{6}$ the 3-tangle, which takes
the form of a hyperdeterminant \cite{Coffman_2000}. Note that $I_{1,...,6}$
are also invariant under a time reversal transformation. Restricted
to the symmetric sector, these invariants explicitly read, with $c_{0}=-v_{12}v_{13}v_{23}$,
$c_{1}=v_{12}v_{13}+v_{12}v_{23+}v_{13}v_{23}$, and $c_{2}=-(v_{12}+v_{13}+v_{23})$, 

\begin{eqnarray}
I_{2,3,4} & = & \frac{-6c_{0}+18c_{1}+\left(c_{2}-60\right)c_{2}+75}{9\left(c_{2}-3\right){}^{2}},\nonumber \\
I_{5} & = & \frac{1}{18\left(c_{2}-3\right){}^{3}}\times\left[-9c_{0}\left(c_{2}-9\right)-459+\right.\nonumber \\
 &  & \left.+27c_{1}\left(c_{2}-5\right)+\left(c_{2}-24\right)c_{2}\left(4c_{2}-21\right)\right],\nonumber \\
I_{6} & = & \frac{2\left(c_{0}+c_{1}+c_{2}+1\right)}{3\left(c_{2}-3\right){}^{2}}.\end{eqnarray}

\begin{figure}[ht]
\begin{centering}
\includegraphics[width=0.9\columnwidth]{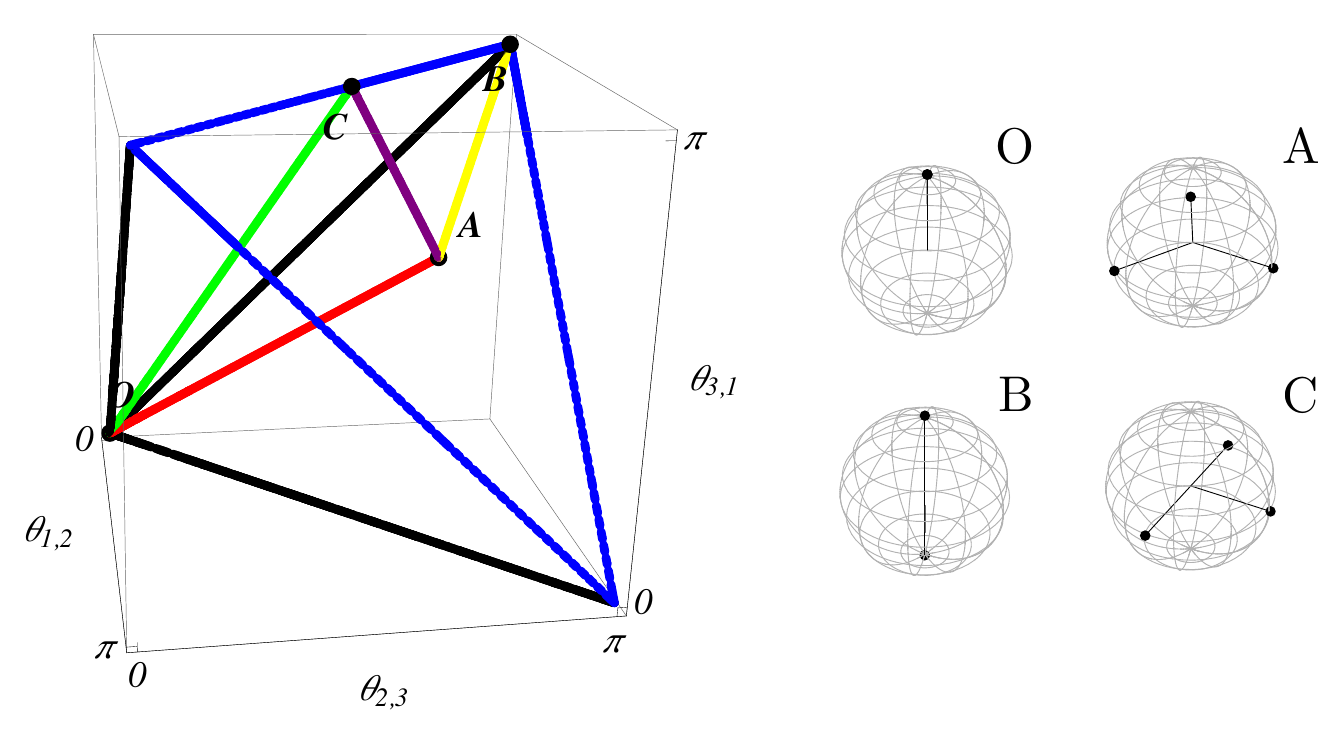} 
\par\end{centering}

\caption{\label{fig:geo_1}Entanglement types, for symmetric $3\mbox{-qubits }$space,
in the $\theta_{i,j}$ space. Point $O$ corresponds to separable
states (with coinciding 3 Majorana points), B and A to $W$ and $GHZ$
states respectively. }

\end{figure}

Using $\theta_{i,j}=\arccos v_{i,j}$ as coordinate axes, and recalling
that the set of Majorana points is not ordered, we can display the
symmetric sector entanglement types inside the tetrahedron ($OABC$)
shown in Fig.\ref{fig:geo_1}. Analyzing the subgroups of $SO(3)$
that leave each representative state invariant one can characterize
the manifold corresponding to each entanglement family (see Table
\ref{tab:geo_3}).

\begin{table}[H]
\centering{}$\begin{array}{ccccc}
\text{States} & \text{Manifold} & I_{2} & I_{5} & I_{6}\\
O & S^{2} & 1 & 1 & 0\\
A & SO(3)/Z_{3} & 0 & 1/4 & 1^{(**)}\\
B & S^{2} & 1/9 & 2/9^{(*)} & 0\\
C & SO(3) & 4/9 & 17/36 & 1/3\end{array}$\caption{\label{tab:geo_3}Manifold of the particular points O, A, B and C
of Fig. \ref{fig:geo_1}. ({*}) $2/9$ is the minimum of $I_{5}$
within the class of symmetric states arising only for type B states.
({*}{*}) Maximal 3-tangle states. }

\end{table}

\begin{center}
\textit{Toward a determination of the unit 3-tangle manifold}
\par\end{center}

Symmetric GHZ states (with unit 3-tangle $I_{6}=1$) correspond to
the three Majorana points forming an equilateral triangle on an equatorial
plane. The set of equatorial planes is the projective plane $RP^{2}$.
Adding the triangles global rotation modulo $2\pi/3$, the set of
symmetric unit 3-tangle states inherits the geometry $SO(3)/Z_{3}$. 

Using the above defined time reversal operator $T$ , we consider
the operator $Y\left(\theta\right)=\left(\cos\theta+\sin\theta T\right)$,
whose inverse is $Y\left(-\theta\right)$ (since $T^{2}=-1$ for $n$
odd); $Y\left(\theta\right)$ is left unchanged under conjugation
with a LU. Applying $Y\left(\theta\right)$ onto a separable 3-qubit
state, one gets interesting entangled states. Starting from a symmetric
separable state, one proves that any symmetric GHZ state can be obtained
as $\ket{\Psi}=Y\left(\frac{\pi}{4}\right)\ket u\ket u\ket u$. More
generically, $Y\left(\frac{\pi}{4}\right)$ maps a non symmetric separable
state $\ket{u_{1}}\ket{u_{2}}\ket{u_{3}}$ onto a (non-symmetric)
GHZ state, as can be verified by a direct check. One can show that
these GHZ states form the manifold $\mathcal{M=\mathrm{S^{2}\times S^{2}\times}\mathrm{SO(3)/Z_{3}}}$.
In the case (yet unproved, but numerically plausible) that any generic
unit 3-tangle GHZ can be sent to the symmetric space by a LU, this
would prove that the full GHZ manifold is indeed $\mathcal{M}$. Note
$\mathcal{M}$ differs by a factor $Z_{3}$ from that given in \cite{Sarbick_2005}.

\begin{center}
\textit{Symmetric states SLOCC invariants}
\par\end{center}

\begin{figure}[ht]
\begin{centering}
\includegraphics[scale=0.6]{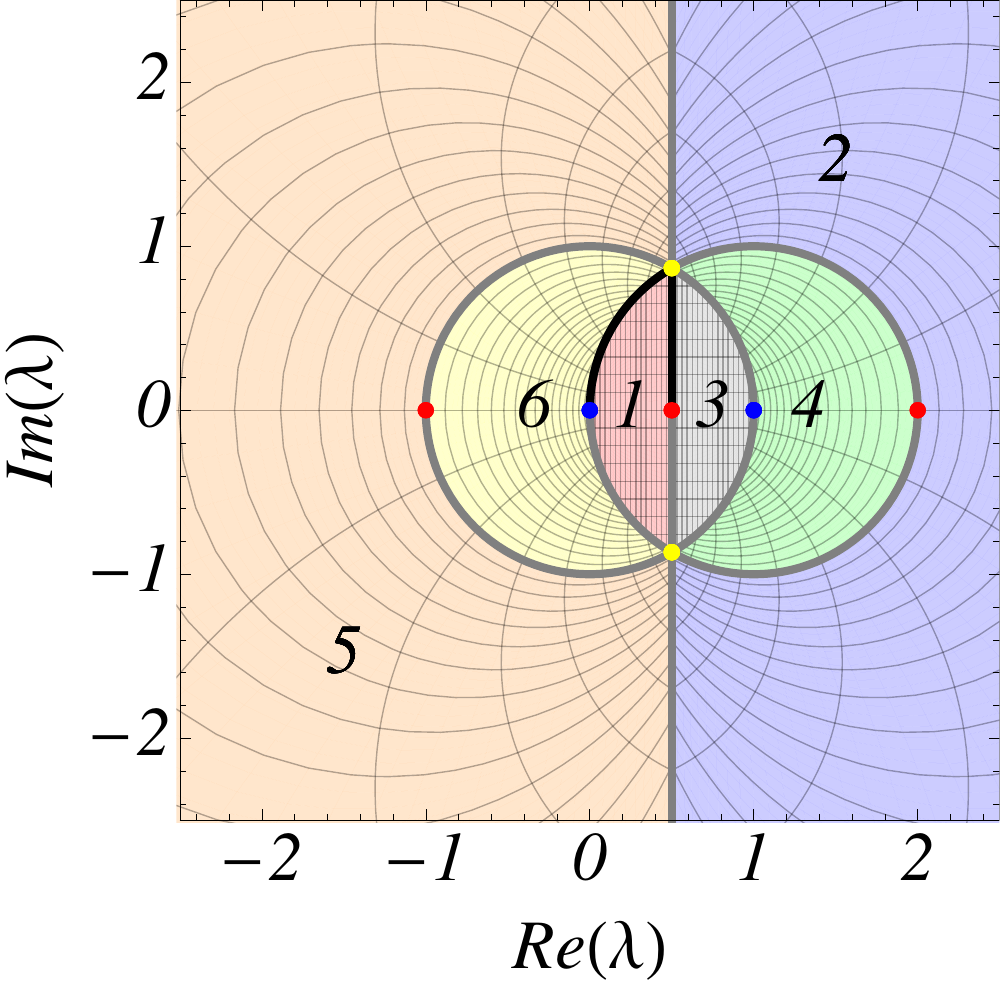}
\par\end{centering}

\caption{\label{fig:domains}Symmetries of the Cross Product. For a given set
of four complex numbers, the 6 permutation related cross ratios belong
to separate regions $\mathcal{D_{\mathit{\mathrm{i}}}}$ labeled from
$1$ to $6$ in the picture. The boundaries of the regions carry more
symmetries, so one should for example only consider the black lines
for region $\mathcal{D_{\mathit{1}}}$. States associated with invariant
on the boundary set, like the colored ones, are expected to display
particular properties.}

\end{figure}

A nice description of SLOCC invariant families was recently proposed
for symmetric $n$-qubits states \cite{Bastin_2009,Mathonet_2010},
which focuses on the number of different roots $\alpha_{i}$ and their
degeneracy. This allows a full classification for $n=2$ or $3$ but,
as stressed by the authors, leaves continuous families of additional
parameters for larger $n$. Our aim here is to provide a closer look
to this question, by mapping this problem to the classification of
invariants of Möbius transformations. Indeed, an ILO $A$ that leaves
the symmetric sector invariant can also be parameterized as in (\ref{eq:sym_trans}),
but now with $\mathbf{h}$ being complex instead of real. Upon simple
manipulations, one parameterize this transformation as\begin{eqnarray}
A & = & e^{ih\left(\frac{1}{\beta_{1}+\beta_{2}}S_{+}+S_{z}-\frac{\beta_{1}\beta_{2}}{\beta_{1}+\beta_{2}}S_{-}\right)},\end{eqnarray}
 where $\beta_{1},\beta_{2},h\in\mathbb{C}$. The action of this operator
on a generic state in the coherent state basis is given by \cite{Perelomov_1986}
\begin{eqnarray}
A\Psi\left(\alpha\right) & = & \left[\frac{\gamma^{-1}\left(\alpha-\beta_{1}\right)-\gamma\left(\alpha-\beta_{2}\right)}{\left(\beta_{1}-\beta_{2}\right)}\right]^{2s}\times\nonumber \\
 &  & \Psi\left(\frac{\gamma^{-1}\beta_{2}\left(\alpha-\beta_{1}\right)-\gamma\beta_{1}\left(\alpha-\beta_{2}\right)}{\gamma^{-1}\left(\alpha-\beta_{1}\right)-\gamma\left(\alpha-\beta_{2}\right)}\right),\end{eqnarray}
where $\gamma=e^{i\frac{h}{2}\frac{\left(\beta_{1}-\beta_{2}\right)}{\left(\beta_{1}+\beta_{2}\right)}}$.
Note that this transformation lets the wave function invariant for
$\alpha=\beta_{1}$ and $\alpha=\beta_{2}$. The roots $\alpha_{i}$
of the polynomial $\Psi\left(\alpha\right)$ transform according to
the following Möbius Transformation (MT): \begin{eqnarray}
\alpha_{i} & \to\alpha'_{i}= & \frac{(\beta_{2}\gamma-\beta_{1}\gamma^{-1})\alpha_{i}+\beta_{1}\beta_{2}(\gamma^{-1}-\gamma)}{(\gamma-\gamma^{-1})\alpha_{i}+\gamma^{-1}\beta_{2}-\gamma\beta_{1}}.\label{eq:root_transf-1}\end{eqnarray}
Unitary transformations are recovered whenever $\beta_{1}=-\bar{\beta}_{2}^{-1}$
and $h\in\mathbb{R}$, corresponding to the sub-class of elliptic
MT. This mapping from ILO to MT is particularly interesting when looking
to invariant quantities. Indeed, the latter are well known to preserve
the {}``cross-ratio'' of four (here complex) numbers: \begin{eqnarray}
(\alpha_{i},\alpha_{j};\alpha_{k},\alpha_{l}) & = & \frac{\left(\alpha_{i}-\alpha_{k}\right)\left(\alpha_{j}-\alpha_{l}\right)}{\left(\alpha_{j}-\alpha_{k}\right)\left(\alpha_{i}-\alpha_{l}\right)}\label{eq:cross_r}\end{eqnarray}
 which therefore form the natural basis for SLOCC invariants. Note
that permuting the roots $\alpha$ in the cross ratio $(\alpha_{i},\alpha_{j};\alpha_{k},\alpha_{l})=\lambda$
leads generically to the following 6 different values for the cross
ratio out of the 24 permutations: $\left\{ \lambda,\frac{1}{\lambda},1-\lambda,\frac{1}{1-\lambda},\frac{\lambda}{\lambda-1},\frac{\lambda-1}{\lambda}\right\} $,
belonging to distinct regions in the complex plane (Fig. \ref{fig:domains}).

As discussed in \cite{Bastin_2009}, for $n$ qubits, the symmetric
SLOCC classes are parameterized by $n-3$ continuous parameters. In
terms of MT, this is nothing but the known property that a unique
MT relate two sets of three distinct complex numbers, and that transformations
involving $n$ complex numbers are parameterized by $n-3$ cross ratios.
This immediately recovers the result that, for $n=3$, there are 3
SLOCC classes in the symmetric sector, labeled by the points $O$,
$B$, and $A$ in Fig.\ref{fig:geo_1}: Separable states (point $O$),
with the three roots $\alpha_{i}$ identical, W states (point $B$)
with two roots identical and the remaining (generic) states that can
be mapped under SLOCC to the GHZ state (point $A$).

A complete set of SLOCC invariants (for any $n$) can be obtained
by choosing $3$ roots $\alpha_{i}$ ($i=1,2,3$) in order to define
the function $\lambda\left(z\right)=\frac{(z-\alpha_{1})(\alpha_{2}-\alpha_{3})}{(z-\alpha_{3})(\alpha_{2}-\alpha_{1})}$.
The $n-3$ complex values $\boldsymbol{\lambda}=\left\{ \lambda_{1},...,\lambda_{n-3}\right\} $,
where $\lambda_{j-3}=\lambda(\alpha_{j})$ for each $\alpha_{j>3}$,
form the SLOCC invariants. Since the ordering of the $n$ roots is
arbitrary, there are in general $n!$ such sets: under a permutation
$\Pi$ the cross ratios transform as $\boldsymbol{\lambda}\to\boldsymbol{\lambda}'\left(\Pi\right)$
where each $\lambda'_{j}\left(\Pi\right)$ is a rational function
of the $\lambda_{j}$'s. 

For $n=4$, we noted the reduction to $6$ independent transformations;
the requirement that $\lambda=\lambda(\alpha_{4})\in\mathcal{D_{\mathit{1}}}$
fixes then a unique value of the SLOCC invariant. In Ref.\cite{Bastin_2009}
a state having four different roots was shown be SLOCC equivalent
to a state within the one-parameter family: $\ket{\Psi\left(\mu\right)}=\ket{\text{GHZ}_{4}}+\mu\ket{D_{4}^{(2)}}$
with $\mu\in\mathbb{C}\backslash\left\{ -\frac{1}{\sqrt{3}},\frac{1}{\sqrt{3}}\right\} $
(where $\ket{\text{GHZ}_{4}}=\frac{1}{\sqrt{2}}\left(\ket{s=2,m=-2}+\ket{s=2,m=2}\right)$
and $\ket{D_{4}^{(2)}}=\ket{s=2,m=0}$). Computing the cross ratio
for this family one obtains the relation $\lambda=\frac{1}{2}\left(\sqrt{3}\mu+1\right)$. 

For $n>4,$ the set of permutation related cross ratios leads to complex
geometrical patterns and the identification of a canonical domain
analogous to $\mathcal{D}_{1}$ is difficult (as an example for $n=5$
all $5!$ transformations leads to inequivalent cross ratio sets).
We therefore introduce a more symmetrical formulation of the invariant
quantities, $I_{k}(\boldsymbol{\lambda})=\sum_{\Pi}\left[\lambda'_{1}\left(\Pi\right)\right]^{k}$,
which amounts to sum the $k^{th}$powers of the transformed cross
ratios (say of $\lambda'_{1}$) over the complete orbit of the permutation
group. Back to $n=4$, a non trivial symmetrized invariant $I_{2}(\lambda)$
is obtained :

\begin{eqnarray*}
I_{2}\left(\lambda\right) & = & \frac{2\left(\lambda^{6}+1\right)-6\left(\lambda^{5}+\lambda\right)+9\left(\lambda^{4}+\lambda^{2}\right)-8\lambda^{3}}{(\lambda-1)^{2}\lambda^{2}}\\
 & = & -3+\frac{27}{2}J(\lambda)\end{eqnarray*}
where $J(\lambda)$ is known as the Klein modular invariant \cite{Klein_1921}.
The next case is $n=5$, where two independent invariants $I_{2}\left(\lambda_{1},\lambda_{2}\right)$
and $I_{4}\left(\lambda_{1},\lambda_{2}\right)$ can be generated
by summing the cross ratios squares and fourth powers over the $120$
permutations. Due to lack of space, the explicit form of the two invariants
is not given here. When two Majorana roots are equal one can, without
loss of generality, let $\lambda_{1}$ go to zero, in which case both
invariants diverge, but we find again the Klein invariant in the following
expression $\lim_{\lambda_{1}\to0}\frac{I_{4}\left(\lambda_{1},\lambda_{2}\right)}{I_{2}\left(\lambda_{1},\lambda_{2}\right)^{2}}=\frac{1}{8}-\frac{2}{27J\left(\lambda_{2}\right)},$
which allow to fully characterize the states having 3 or 4 different
roots.

In conclusion we have explicitly constructed a set of entanglement
invariants under LOCC and SLOCC for symmetric $n$-qubit states and
given several examples for $n$ up to five. We also expect that this
correspondence between ILO and Moëbius Transformations, may find further
possible experimental consequences. Indeed, a generic Möbius transform
can be decomposed into elementary operations, such as translations,
rotations, inversions and dilation. It would therefore be very interesting
to perform such elementary operations by implementing suitable POVM's
within the symmetric sector. 
\begin{acknowledgments}
PR acknowledges support through FCT BPD grant SFRH/BPD/43400/2008.
RM also acknowledges discussions with M. Kus and K.Zyczkowski about
the unit 3-tangle manifold and with J-M Maillard, about SLOCC invariants,
who in particular noted the connection between $I\left(\lambda\right)$
and the Klein invariant.
\end{acknowledgments}

\end{document}